# Dynamic Pricing of Power in Smart-Grid Networks

Qingsi Wang, Mingyan Liu and Rahul Jain

## Abstract

In this paper we introduce the problem of dynamic pricing of power for smart-grid networks. This is studied within a network utility maximization (NUM) framework in a deterministic setting with a single provider, multiple users and a finite horizon. The provider produces power or buys power in a (deterministic) spot market, and determines a dynamic price to charge the users. The users then adjust their demand in response to the time-varying prices. This is typically categorized as the demand response problem, and we study a progression of related models by focusing on two aspects: 1) the characterization of the structure of the optimal dynamic prices in the Smart Grid and the optimal demand and supply under various interaction with a spot market; 2) a greedy approach to facilitate the solution process of the aggregate NUM problem and the optimality gap between the greedy solution and the optimal one.

## I. INTRODUCTION

As the Smart Grid takes shape, new possibilities of efficient management of the electric power grid open up. One of these is pricing of electricity to consumers. Currently, temporal variations in the cost of electricity are hidden behind inflexible rate designs. This leads to inefficiencies with over-consumption during peak times, and under-consumption during off-peak times. It also makes the problem of matching demand and supply, both of which are uncertain, and affect grid stability, particularly acute. In the United States, only 60-100 hours of the year can account for 10-18 percent of the system peak load [1]. Meeting this critical peak load requires installation, operation and maintenance of expensive combustion gas-turbine generators, since these start almost instantaneously.

Dynamic pricing of electricity to consumers can remedy this problem by inducing consumers to switch off or defer some of the non-urgent, non-critical loads. For example, dish-washers in most households have a delayed start option. And yet, only a small fraction of consumers utilize this feature and use the appliance later at night at off-peak times. Dynamic pricing can make consumers sensitive to their time and amount of electricity consumption, thus potentially smoothing out peak-time system loads, and enhancing economic efficiency.

Dynamic pricing rate designs are receiving increased attention by state commissions, with the California Public Utilities Commission (CPUC) having set a deadline of 2011 for the state electric utilities to propose dynamic pricing rate structures. These are defined as an electric rate structure that reflects the actual wholesale market conditions. Dynamic pricing can take various forms such as *critical peak-pricing*, *time-of-use pricing* and *real-time pricing*. CPUC defines the real-time price as the rate linked to the actual price in the wholesale hourly electricity market [2]. Such prices can be communicated to the consumers over the *advanced metering infrastructure* (AMI), or a *smart-grid network* [3] and displayed on a *smart meter*.

*Literature Overview.* Problems related to the design and operation of a Smart Grid are receiving increasing attention. The problem of dynamic pricing to shape demand is typically called *demand response*. There exists

Q. Wang and M. Liu are with the Department of Electrical Engineering and Computer Science, University of Michigan, Ann Arbor, MI 48109-2122, USA. Email: (qingsi, mingyan)@umich.edu

R. Jain is with the Department of Electrical Engineering and the Department of Industrial and Systems Engineering, University of Southern California, Los Angeles, USA. Email: rahul.jain@usc.edu

a large literature on demand response; the most relevant is [4]. This paper formulates the problem of optimal demand response in discrete time over a finite time horizon when users have some storage available, and the formulation can be categorized as a network utility maximization (NUM) problem (see [5], [6] and the references therein). Using Lagrangian duality, it establishes the existence of a solution, and gives a distributed algorithm based on gradient projection to compute this solution. The modeling of various appliances into the proposed optimization framework is also studied. In [7], a different approach is taken wherein demand is required to match supply, and users bid for load-shedding in an iterative supply function bidding mechanism, that was first proposed in [8]. Both competitive and Nash equilibrium analysis is provided but the focus is on a single instant, and without storage. In [9], the focus is on optimal power flow through the network, wherein existence of a zero duality gap solution is proved. In contrast to aforementioned work, particularly [4], in this paper we study a progression of related models by focusing on two aspects: 1) the characterization of the structure of the optimal dynamic prices in the Smart Grid and the optimal demand and supply under various interaction with a spot market; 2) a greedy approach to facilitate the solution process of the aggregate NUM and the optimality gap between the greedy solution and the optimal one.

Other works have focused on electricity markets. While [10] provides a stability analysis of the wholesale electricity market, [11] has focused on means for wind power generators to participate in the day-ahead electricity market. The difficulty in this is that wind-power generators cannot enter into binding contracts for a day-ahead power supply without assuming a huge risk. Thus, alternative market architectures are considered in [12], wherein entities called "aggregators" buy power from the wind power generators and participate in the day-ahead markets. The paper proposes optimal mechanisms that the aggregators can use to get the wind power generators to reveal their true distributions, and thus minimize the risk due to uncertainty in supply from the contracted wind power generators. A related paper is [13], wherein mechanisms for riskless dispatch by the generators are given for use with the current market architecture.

*Results and organization of this paper*. The system model is first presented in Section II, and three related problems are introduced. The first problem SYSTEM is a simplified special case of the main model with internalized production cost, which provides a preliminary framework for more realistic refinement. Its optimal solution is characterized in Section III based on a dual decomposition; moreover, a distributed algorithm to solve SYSTEM, which can be readily adapted for problems in the same family, is also proposed. In Section IV we extend the SYSTEM problem to the scenario where the provider may purchase in part from a spot market with an exogenous cost to narrow the demand-supply gap. This is referred to as the SYS_SPOT problem, and the interaction between the decision of purchase and the spot market price is investigated. We then consider the explicit effect of the production cost in a third problem in Section V, and the structure of its optimal control in the single-user scenario is studied. A greedy algorithm to approximate the solution to SYSTEM is presented in Section VI, where an upper bound on the optimality gap is established and heuristics on evaluating this gap are discussed. Numerical results of sample problems are illustrated in Section VII. We present an attempt of adopting distributed Newton's method to problems in this work in Section VIII, and Section IX concludes the paper.

## II. SYSTEM MODEL AND NOTATION

Consider a smart-grid network that consists of $n$ users or households, and one provider. We consider a discrete-time model with a finite horizon $t_f$. A user $i$ consumes $x_i(t)$ power at time step $t$, and the supply by the grid is denoted as $z_i(t)$. The excess supply/demand is charged to/discharged from the user's battery with capacity $B_i$. The energy level/state of the battery before operation at time $t$ is denoted by $y_i(t)$. We assume that the total supply from the grid at time $t$ is upper bounded by $C_t$, and the total demand of user $i$ over time is upper bounded by $D_i$. For convenience we define the following vector forms of the above quantities: the demand of user $i$, $x_i = \begin{bmatrix} x_i(0), & x_i(1), & \cdots, & x_i(t_f) \end{bmatrix}^T$; the supply of the grid $z_i = \begin{bmatrix} z_i(0), & z_i(1), & \cdots, & z_i(t_f) \end{bmatrix}^T$; the state of user $i$'s battery, $y_i = \begin{bmatrix} y_i(1), & y_i(2), & \cdots, & y_i(t_f+1) \end{bmatrix}^T$; the battery capacities of users $B = \begin{bmatrix} B_1, & B_2, & \cdots, & B_n \end{bmatrix}^T$; the supply constraint of the grid $C = \begin{bmatrix} C_0, & C_1, & \cdots, & C_{t_f} \end{bmatrix}^T$; the demand constraint of users $D = \begin{bmatrix} D_1, & D_2, & \cdots, & D_n \end{bmatrix}^T$. Also, $x = \text{vect}(x_1, \ldots, x_n)$ with $\text{vect}(u_1, \ldots, u_n)$ being the stacked vector from $u_1$ to $u_n$, and similarly define $y$ and $z$. When user $i$ draws power $x_i$ over time, her satisfaction is measured by the utility function

$$U_i(x_i) = \sum_{t=0}^{t_f} V_i(x_i(t)),$$

where $V_i : \mathbf{R}_+ \to \mathbf{R}_+$ is a strictly increasing, strictly concave and twice differentiable function of $x_i(t)$. Denote by $\dot{y}_i(t)$ the change of battery state of user $i$ at time $t$:

$$\dot{y}_i(t) = y_i(t+1) - y_i(t) = z_i(t) - x_i(t).$$

Assuming zero initial state, $y(0) = 0$, we then have

$$y_i(t) = y_i(0) + \sum_{s=0}^{t-1} \dot{y}_i(s) = \sum_{s=0}^{t-1} (z_i(s) - x_i(s)).$$

Therefore,

$$y_i = H(z_i - x_i),$$

where

$$H = \begin{bmatrix} 1 & 0 & \cdots & 0 & 0 \\ 1 & 1 & \ddots & 0 & 0 \\ \vdots & & \ddots & \ddots & \vdots \\ 1 & 1 & \cdots & 1 & 0 \\ 1 & 1 & \cdots & 1 & 1 \end{bmatrix} \in \mathbf{R}^{(t_f+1)\times(t_f+1)}.$$

In addition, since the battery suffers wear and tear due to usage, we model maintenance and operational costs as

$$M(y_i) = \sum_{t=1}^{t_f+1} m(y_i(t)),$$

where $m : \mathbf{R}_+ \to \mathbf{R}_+$ is a strictly increasing, strictly convex and twice differentiable function of $y_i(t)$. The objective is to maximize the sum of individual utilities of both users and the provider or the social welfare. Formally, we consider the following related problems in this paper. The main model is given as follows.

$$\text{maximize} \quad \sum_{i=1}^{n} \big(U_i(x_i) - M(y_i)\big) - P(z) \tag{1}$$

$$\text{subject to} \quad y_i \preceq B_i \mathbf{1} \tag{2}$$

$$y_i = H(z_i - x_i) \tag{3}$$

$$\sum_{i=1}^{n} z_i \preceq C \tag{4}$$

$$\mathbf{1}^T x_i \leq D_i \tag{5}$$

$$x_i, y_i, z_i \succeq 0, i = 1, 2, \ldots, n \tag{6}$$

where $P : \mathbf{R}_+^{n(t_f+1)} \to \mathbf{R}_+$ is the production cost function to be specified later, and $\mathbf{1} = (1, 1, \ldots, 1)^T$ with proper sizes. We begin with a simplified special case of this problem by assuming $P = 0$; this is denoted as SYSTEM $(U, M, B, H, C, D)$. The motivation for studying this degenerated model is twofold: 1) it models the scenario when the production cost is realized thus fixed *before* the distribution of energy or when the marginal cost is zero such as in the case of renewable energies; 2) more critically, it provides a preliminary framework that reveals the basic structure of optimal solutions and their algorithmic implementation that can be readily extended to more realistic models.

We then consider the setting wherein the grid may purchase power from an external spot market to narrow the demand-supply gap, while incurring an exogenous cost; this is formulated as the SYS_SPOT $(U, M, B, H, C, D; \sigma)$ problem as follows.

$$\text{maximize} \quad \sum_{i=1}^{n} \left( U_i(x_i) - M(y_i) \right) - \sigma^T \left[ \sum_{i=1}^{n} z_i - C \right]^+ \tag{7}$$

$$\text{subject to} \quad y_i \preceq B_i \mathbf{1} \tag{8}$$

$$y_i = H(z_i - x_i) \tag{9}$$

$$\mathbf{1}^T x_i \leq D_i \tag{10}$$

$$x_i, y_i, z_i \succeq 0, i = 1, 2, \ldots, n \tag{11}$$

where $\sigma = \begin{bmatrix} \sigma_0 & \sigma_1 & \ldots & \sigma_{t_f} \end{bmatrix}$ is the price vector of unit power in the spot market, and $[\cdot]^+ = \max\{\cdot, 0\}$. Our main interest in this problem is the characterization of spot market prices that may incentivize the exogenous purchase. In particular, we show that there is a threshold price below which this purchase is justified, and it is closely related to the optimal solution to SYSTEM.

Our last model considers non-trivial production costs. We consider the linear cost in the original setting with multiple users, that is, $P(z) = \sigma^T d_z$, where $d_z = \sum_{i=1}^{n} z_i$ is the vector of aggregate demands, and $\sigma$ can be either interpreted as the unit power generation cost or the spot market price when the provider produces nothing of its own and completely resorts to the spot market for supply. This model is denoted by SYS2 $(U, M, B, H, C, D)$. We also consider a general convex cost in the single-user case that has an explicit control structure.

In the next section we start with the SYSTEM problem, and note that when constraint (4) is decoupled, the SYSTEM problem can be separated into several subproblems. This observation motivates us to consider the dual-decomposition based technique to solve the SYSTEM problem [5], [6], which we show next.

### III. Dual Decomposition of SYSTEM and the Properties of Its Solutions

In this section the SYSTEM problem is analyzed based on a dual decomposition, which also prompts a distributed first-order algorithm to solve it, and the properties of both the primal and the dual optimal solutions (demand, supply and prices) are discussed.

## A. Dual decomposition

We first reformulate SYSTEM $(U, M, B, H, C, D)$ as

$$\text{maximize} \quad \sum_{i=1}^{n} \left( U_i(x_i) - M(y_i) \right) \tag{12}$$

$$\text{subject to} \quad \text{vect}(x, y, z) \in \mathcal{F} \tag{13}$$

$$\text{over} \quad \text{vect}(x, y, z) \in \mathcal{D} \tag{14}$$

where the domain and the feasible region are respectively given by

$$\mathcal{D} = \left\{ \text{vect}(x, y, z) \; \middle| \; \begin{array}{l} x_i \succeq 0, \mathbf{1}^T x_i \leq D_i, \\ 0 \preceq y_i \preceq B_i \mathbf{1}, z_i \succeq 0, \forall i \end{array} \right\},$$

and

$$\mathcal{F} = \left\{ \text{vect}(x, y, z) \; \middle| \; y_i = H(z_i - x_i), \sum z_i \preceq C, \forall i \right\}.$$

Note that **relint** $\mathcal{D} \cap \mathcal{F} \neq \emptyset$ when $B$, $C$ and $D$ are all positive, as we shall assume so. Indeed, let $\epsilon > 0$ be chosen later, and consider

$$\hat{x}_i(t) = \alpha \frac{D_i - \epsilon}{t_f + 1},$$

where $0 < \alpha \leq 1$ satisfying

$$\alpha \frac{\sum_{i=1}^{n} D_i}{t_f + 1} < \min_{t} C_t,$$

$\hat{z}_i(t) = \hat{x}_i(t) + \frac{\alpha \epsilon}{t_f + 1}$ and $\hat{y}_i(t) = \sum_{s=0}^{t-1} (\hat{z}_i(s) - \hat{x}_i(s))$. Let $\epsilon < \min_i \{\min\{B_i, D_i\}\}$, and it can be verified that $\text{vect}(\hat{x}, \hat{y}, \hat{z}) \in \textbf{relint}\,\mathcal{D} \cap \mathcal{F}$. In addition, since this problem is convex and the inequality constraint is affine, Slater's condition is then satisfied, and hence, strong duality will hold. Because of the strict concavity of the objective function in $x$ and $y$, there are unique optima for $x$, $y$ and $z$. As pointed out before, the structure of the SYSTEM problem prompts a solution based on a dual decomposition. Its Lagrangian is given by

$$L = \sum_{i=1}^{n} \left( U_i(x_i) - M(y_i) \right) + \sum_{i=1}^{n} \nu_i^T \left( H(z_i - x_i) - y_i \right) + \theta^T \left( C - \sum_{i=1}^{n} z_i \right)$$

$$= \sum_{i=1}^{n} \left( U_i(x_i) - \nu_i^T H x_i \right) - \sum_{i=1}^{n} \left( M(y_i) + \nu_i^T y_i \right) - \sum_{i=1}^{n} (\theta - H^T \nu_i)^T z_i + \theta^T C.$$

Let $\nu = \text{vect}(\nu_1, \ldots, \nu_n)$. The dual function is then given by

$$g(\nu, \theta) = \max_{\text{vect}(x,y,z) \in \mathcal{D}} L(x, y, z, \nu, \theta)$$

$$= \sum_{i=1}^{n} \max_{\substack{x_i \succeq 0, \\ \mathbf{1}^T x_i \leq D_i}} W_i(x_i; \nu_i) - \sum_{i=1}^{n} \min_{0 \preceq y_i \preceq B_i \mathbf{1}} R_i(y_i; \nu_i) - \sum_{i=1}^{n} \min_{z_i \succeq 0} (\theta - H^T \nu_i)^T z_i + \theta^T C,$$

where

$$W_i(x_i; \nu_i) = U_i(x_i) - \nu_i^T H x_i,$$

and

$$R_i(y_i; \nu_i) = M(y_i) + \nu_i^T y_i,$$

and we introduce

$$\texttt{USER}_\texttt{i} : \text{maximize} \quad W_i(x_i; \nu_i)$$
$$\text{subject to} \quad x_i \succeq 0, \mathbf{1}^T x_i \leq D_i \tag{15}$$

and

$$\texttt{USER}'_\texttt{i} : \text{minimize} \quad R_i(y_i; \nu_i)$$
$$\text{subject to} \quad 0 \preceq y_i \preceq B_i \mathbf{1} \tag{16}$$

as two subproblems for user $i$. Since the minimization with respect to $z_i$ is a linear problem, which is readily solved, the dual problem is given by $\texttt{DUAL\_SYS}\ (U, M, B, H, C, D)$:

$$\text{minimize} \quad \sum_{i=1}^{n} \left( W_i^\star(\nu_i) - R_i^\star(\nu_i) \right) + \theta^T C$$
$$\text{subject to} \quad \theta - H^T \nu_i \succeq 0, \theta \succeq 0, i = 1, 2, \ldots, n$$

where $W_i^\star(\nu_i)$ and $R_i^\star(\nu_i)$ are the optimal values of problem (15) and (16) for a given $\nu_i$. Given now this dual decomposition structure, a distributive algorithm can be readily proposed to solve SYSTEM, and this is widely studied in the literature, see e.g. [5], [4]. We report our version of the algorithm in Appendix A and show its convergence.

### B. Properties of solutions to SYSTEM and DUAL_SYS

Denote by $x^\star = \text{vect}(x_1^\star, \ldots, x_n^\star)$ the optimum for $x$ in SYSTEM; $\nu^\star$ is an optimal solution for $\nu$ of DUAL_SYS, and similarly define $y^\star$, $z^\star$ and $\theta^\star$, respectively. We first note that $H^T \nu_i$ can be interpreted as the price vector that the provider charges from user $i$ for each unit of consumed energy, in light of subproblem $\texttt{USER}_\texttt{i}$, given that it is component-wise non-negative. Indeed, the following result establishes this premise.

*Proposition 1:* $H^T \nu_i^\star \succeq 0$ for all $i$.

*Proof:* The proof can be found in Appendix B. ∎

*Remark 1:* Since $z_i(t)$ is the total amount of energy distributed to user $i$ at time $t$, under the interpretation of $H^T \nu_i$ being a price vector it is as if the energy buffered in battery were only charged when it is later consumed while incurring a *storage cost before* the consumption. This obviously is not what actually happens in practice, but serves as an interesting interpretation of the result. We do note that if there is no battery buffering in the model, this interpretation is consistent with the notion of shadow price in the literature (see e.g. [14], [15]), i.e., entries of $H^T \nu_i$ are shadow prices of power.

Hence, $\theta^\star(t) = \max_i \left( H^T \nu_i^\star \right)_t$ for all $t$. Moreover, 1) if we assume that $x_i^\star \in \textbf{int}\,\mathcal{X}_i$ for all $i$, where $\mathcal{X}_i$ is the feasible set of (15), i.e., $\mathcal{X}_i = \{ x_i \mid x_i \succeq 0, \mathbf{1}^T x_i \leq D_i \}$, $H^T \nu_i^\star$ is then uniquely determined for each $i$; 2) alternatively, if we assume that $x_i^\star \in \textbf{int}\,\mathcal{X}_i$ for at least one $i$, say $i_k$, and $z_{i_k}^\star \succ 0$, $H^T \nu_{i_k}^\star$ is then uniquely determined and from the minimization with respect to $z_{i_k}$ we get that $\theta^\star = H^T \nu_{i_k}^\star$. In either case, $\theta^\star$ is also uniquely determined.

In Section IV, we will consider the situation where there is a gap between the users demand and the power generation capacity of the grid at some time instants. At those instants, the grid may purchase power from the spot market to narrow the demand-supply gap. We will show therein that the above unique $\theta^\star$ is a "lower bound" on the price of unit power in the spot market above which the grid decides to make no purchase.

Also, we can determine the closed form of $y^\star$ given $\nu^\star$ based on (16). Since the objective and the constraint are both separable in time, we conclude

$$y_i^\star(t) = [m'^{-1}(-\nu_i^\star(t-1))]_0^{B_i}$$

for all $i$ and $t > 0$, where $[\cdot]_a^b$ is the projection operator onto $[a,b]$, i.e., $[x]_a^b := \min\{\max\{x,a\},b\}$. Let $\rho_i^\star := H^T \nu_i^\star$, and $\dot\rho_i^\star(t) := \rho_i^\star(t) - \rho_i^\star(t-1) = -\nu_i^\star(t-1)$. We then obtain

$$y_i^\star(t) = [m'^{-1}(\dot\rho_i^\star(t))]_0^{B_i}.$$

If we assume that $0 < y_i^\star(t) < B_i$, the above result can be rewritten as

$$\dot\rho_i^\star(t) = m'(y_i^\star(t)).$$

Therefore, at optimality the marginal cost of reallocating the power purchase from $t-1$ to $t$ is equal to the marginal cost of storing the energy in battery after purchasing in $t-1$, which is intuitively appealing. When the demand is known and fixed a priori, some structural results on the optimal power generation and battery scheduling in a single user setting can be found in [16]. In Section V, we will also discuss in detail the explicit structure of optimal control in the single-user problem when a non-zero production cost is incorporated in our main model.

## IV. SPOT MARKET

In this section, we consider the SYS_SPOT problem. As before, we reformulate this problem and its domain and the feasible region are the same as in the SYSTEM problem. Also, there are unique optima for $x$, $y$ and $z$ because of the strict concavity of the objective function in $x$ and $y$.

Denote by $z^\star(\sigma) = (z_1^\star(\sigma), \ldots, z_n^\star(\sigma))$ the optimal solution for $z$ of SYS_SPOT given $\sigma$. Define

$$\mathcal{S} = \left\{ \sigma \;\middle|\; \sum_{i=1}^n z_i^\star(\sigma) \preceq C \right\}.$$

In the following, we characterize $\mathcal{S}$ using the solutions of the SYSTEM problem and its dual problem DUAL_SYS. To avoid ambiguity, symbols without hat are variables in SYS_SPOT, and others are for the SYSTEM problem and its dual. We assume that either of the two cases below is true:

*Assumption 1:*
a) $x_i^\star \in \mathbf{int}\, \mathcal{X}_i$ for all $i$, or
b) $x_i^\star \in \mathbf{int}\, \mathcal{X}_i$ for at least one $i$, say $i_k$, and $z_{i_k}^\star \succ 0$.

Recall that $x_i^\star$ denotes the optimal solution to the SYS_SPOT problem for $x_i$, and $\mathcal{X}_i$ is the feasible set of (15). In the following proposition, we show that $\mathcal{S}$ can be characterized with a threshold of prices.

*Proposition 2:* $\sigma \in \mathcal{S}$ if and only if $\sigma \succeq \widehat{\theta}^\star$, noting that $\widehat{\theta}^\star$ is the optimal dual variable in DUAL_SYS $(U, M, B, H, C, D)$.

*Proof:* 1) Necessity. First of all, the Lagrangian of SYS_SPOT is given by $L(x, y, z, \nu, \mu)$,

$$L = \sum_{i=1}^n (U_i(x_i) - M(y_i)) - \sigma^T \left[\sum_{i=1}^n z_i - C\right]^+ + \sum_{i=1}^n \nu_i^T \left(H(z_i - x_i) - y_i\right)$$

$$= \sum_{i=1}^n \left(U_i(x_i) - \nu_i^T H x_i\right) - \sum_{i=1}^n \left(M(y_i) + \nu_i^T y_i\right) - \sigma^T \left[\sum_{i=1}^n z_i - C\right]^+ + \sum_{i=1}^n \nu_i^T H z_i.$$

Group the terms in a similar way as in `DUAL_SYS`, and the dual problem is

$$\text{minimize} \quad \sum_{i=1}^{n} \left( W_i^\star(\nu_i) - R_i^\star(\nu_i) \right) - G^\star(\nu; \sigma)$$
$$\text{subject to} \quad \mu_i + \nu_i \succeq 0$$
$$\mu_i \succeq 0, i = 1, 2, \ldots, n$$

where $W_i^\star$ and $R_i^\star$ are the respective optimal values of problems (15) and (16) for a given $\nu_i$ as in `DUAL_SYS`, and $G^\star(\nu; \sigma)$ is the optimal value of

$$\text{minimize} \quad \sigma^T \left[ \sum_{i=1}^{n} z_i - C \right]^+ - \sum_{i=1}^{n} \nu_i^T H z_i \quad (17)$$
$$\text{subject to} \quad z_i \succeq 0, i = 1, 2, \ldots, n$$

Because of strong duality[1], the optimal solution $(x^\star, y^\star, z^\star)$ to `SYS_SPOT` is also the maximizer of the Lagrangian $L(x, y, z, \nu^\star)$ where $\nu^\star$ is an optimal solution of the dual. Therefore, $z^\star$ is an optimal solution of (17) when $\nu = \nu^\star$. Since $\sum_{i=1}^{n} z_i^\star$ is finite, we must have $\sigma \succeq H^T \nu_i^\star$ for all $i$. On the other hand, provided $\sum_{i=1}^{n} z_i^\star \preceq C$, we can obtain the same solution $(x^\star, y^\star, z^\star)$ of `SYS_SPOT` by solving the `SYSTEM` problem with the same parameter set $(U, M, B, H, C, D)$. Recalling the properties of solutions to `SYSTEM` in Section III-B, under assumption a) we have $\nu_i^\star$ and $\widehat{\nu}_i^\star$ are uniquely determined and $\nu_i^\star = \widehat{\nu}_i^\star$ for all $i$. In addition, $\widehat{\theta}^\star(t) = \max_i \left( H^T \widehat{\nu}_i^\star \right)_t$ for all $t$. Under assumption b) we have $\nu_{i_k}^\star = \widehat{\nu}_{i_k}^\star$ and $\widehat{\theta}^\star = H^T \widehat{\nu}_{i_k}^\star$. In either case, it follows that $\sigma \succeq \widehat{\theta}^\star$.

2) *Sufficiency*. We only provide an outline of the proof here. Our strategy is to show that the optimal solution $(\widehat{x}^\star, \widehat{y}^\star, \widehat{z}^\star)$ to the `SYSTEM` problem with optimal dual variables $(\widehat{\nu}^\star, \widehat{\theta}^\star)$ also satisfies the KKT conditions of `SYS_SPOT` when $\sigma \succeq \widehat{\theta}^\star$, and then because of strong duality and convexity of `SYS_SPOT`, it is also the optimal solution of `SYS_SPOT` with optimal dual variables $\nu^\star = \widehat{\nu}^\star$, given the spot market price $\sigma$. To this end, note that the KKT conditions of `SYS_SPOT` only differ from those of `SYSTEM` in the dual feasibility and the stationarity conditions because of the replacement of $\widehat{\theta}$ with $\sigma$ and the non-differentiable term in the objective function. Recall that $\widehat{\theta}^\star(t) = \max_i \left( H^T \widehat{\nu}_i^\star \right)_t$ for all $t$, and we have $\sigma \succeq H^T \widehat{\nu}_i^\star$ if $\sigma \succeq \widehat{\theta}^\star$. Let $\partial_{z_i} L$ denote the subdifferential of the Lagrangian of `SYS_SPOT` with respect to $z_i$ for all $i$. Hence, when $\sigma \succeq \widehat{\theta}^\star$ and $\nu_i = \widehat{\nu}_i^\star$, $0 \in \partial_{z_i} L$ for all $z_i = \widehat{z}_i^\star$, and hence combining other optimality conditions, we obtain $(\widehat{x}^\star, \widehat{y}^\star, \widehat{z}^\star, \widehat{\nu}^\star)$ is a primal-dual optimal solution given $\sigma \succeq \widehat{\theta}^\star$. ∎

## V. `SYS2` AND THE STRUCTURE OF OPTIMAL CONTROL FOR A SINGLE USER

Under some circumstances, we have an explicit form of the solution to the `SYS2` problem, which is given by the following proposition.

*Proposition 3:* At time $t$, $\sum_{i=1}^{n} z_i^\star(t) = C_t$ if $(H^T \nu_i^\star)_t > \sigma_t$ for some $i$, and $z_i^\star(t) = 0$ if $(H^T \nu_i^\star)_t < \sigma_t$.

*Proof:* The results follow from the KKT conditions. Firstly, the dual feasibility and the complementary slackness specify that

$$\theta \succeq 0, \theta^T \left( C - \sum_{i=1}^{n} z_i \right) = 0.$$

---

[1]It can be shown that the reformulated `SYS_SPOT` problem is convex and Slater's condition is satisfied.

Furthermore, consider the Lagrangian of SYS2 and by the stationarity condition we have

$$\sigma + \theta - H^T \nu_i \succeq 0, (\sigma + \theta - H^T \nu_i)^T z_i = 0.$$

for all $i$. Thus, if $(H^T \nu_i)_t > \sigma_t$ for some $i$ at time $t$, then $\theta(t) > 0$. Therefore, $\sum_{i=1}^{n} z_i(t) = C_t$. On the other hand, if $(H^T \nu_i)_t < \sigma_t$, we have $\sigma_t + \theta(t) - (H^T \nu_i)_t > 0$, and hence $z_i(t) = 0$. ∎

In the rest of this section, we focus on the decision-making by a single user and a single provider, and derive the structure of optimal policy. We relax our previous constraints such that they are now separable in time, and we assume that $y(t+1) \in [0, B]$, $x(t) \in [0, D]$ and $z(t) \in [0, C]$ for all $t = 0, \cdots, t_f$, where the user index $i$ is now omitted. We assume that if a quantity $z(t)$ is produced at time $t$, the cost is $P_t(z(t))$ where $P_t$ is strictly increasing and convex in $z(t)$. Most of our notation is consistent with previous sections except that $M$ will signify the scalar maintenance cost. Then, the decision-making problem is given by the following optimal control problem, OPT_CONTROL:

$$\max \sum_{t=0}^{t_f} [V(x(t)) - P_t(z(t)) - M(y(t+1))] \tag{18}$$

$$\text{s.t.} \quad \dot{y}(t) = z(t) - x(t), \forall t \quad \}\rho(t)$$

$$\mu_0(t)\{ \quad 0 \leq y(t) \leq B, \forall t, \quad \}\mu_1(t)$$

$$\lambda_0(t)\{ \quad 0 \leq x(t) \leq D, \forall t, \quad \}\lambda_1(t)$$

$$\theta_0(t)\{ \quad 0 \leq z(t) \leq C, \forall t, \quad \}\theta_1(t)$$

$$y(0) = y_0,$$

where $\dot{y}(t) = y(t+1) - y(t)$. Denote the dual variables to the above constraints by $\rho(t), \mu_0(t), \mu_1(t), \lambda_0(t), \lambda_1(t), \theta_0(t)$ and $\theta_1(t)$, as shown above. Then, due to strong duality, the KKT conditions suggest that there exist $y^\star$, $x^\star$, $z^\star$, $\rho^\star$, $\mu_0^\star$, $\mu_1^\star$, $\lambda_0^\star$, $\lambda_1^\star$, $\theta_0^\star$, $\theta_1^\star$ that satisfy

$$V'(x^\star(t)) - \rho^\star(t) - \lambda_1^\star(t) + \lambda_0^\star(t) = 0, \tag{19}$$

$$-P_t'(z^\star(t)) + \rho^\star(t) - \theta_1^\star(t) + \theta_0^\star(t) = 0, \tag{20}$$

$$-M'(y^\star(t)) + \dot{\rho}^\star(t) - \mu_1^\star(t) + \mu_0^\star(t) = 0, \tag{21}$$

$$\mu_0^\star(t) y^\star(t) = 0, \quad \mu_1^\star(t)[B - y^\star(t)] = 0, \tag{22}$$

$$\lambda_0^\star(t) x^\star(t) = 0, \quad \lambda_1^\star(t)[D - x^\star(t)] = 0, \tag{23}$$

$$\theta_0^\star(t) z^\star(t) = 0, \quad \theta_1^\star(t)[C - z^\star(t)] = 0, \tag{24}$$

such that $(x^\star, z^\star)$ is the optimal control law and $y^\star$ is the state of the system, where $\dot{\rho}(t) = \rho(t) - \rho(t-1)$ for $t \geq 1$.

Now, from KKT conditions (21), we have

$$\rho^\star(t) = \sum_{s=1}^{t} M'(y^\star(s)) + \sum_{s=1}^{t} [\mu_1^\star(s) - \mu_0^\star(s)] + \rho^\star(0). \tag{25}$$

Further, from KKT conditions (20), we have

$$P_t(z^\star(t)) = \rho^\star(t) - [\theta_1^\star(t) - \theta_0^\star(t)]$$

for all $t$. From this, using the complementary slackness conditions (24), we can conclude that either $P_t(0) = \rho^\star(t) + \theta_0^\star(t)$, or $P_t(C) = \rho^\star(t) - \theta_1^\star(t)$, or $P_t(z^\star(t)) = \rho^\star(t)$. We can summarize this as

$$z^\star(t) = \left[P_t^{-1}(\rho^\star(t))\right]_0^C. \tag{26}$$

Similarly, from KKT conditions (19), we get

$$V'(x^\star(t)) = \rho^\star(t) + (\lambda_1^\star(t) - \lambda_0^\star(t)),$$

which together with the complementary slackness conditions (23), implies that

$$x^\star(t) = \left[V'^{-1}(\rho^\star(t))\right]_0^D. \tag{27}$$

From (26) and (27), we conclude that $z^\star(t)$ is non-decreasing and $x^\star(t)$ is non-increasing in $\rho^\star(t)$ for each $t$. Moreover, $x^\star(t+1) \leq x^\star(t)$ whenever $\rho^\star(t+1) \geq \rho^\star(t)$. In the following, we assume that $P_t = P$, and consequently $z^\star(t+1) \geq z^\star(t)$ whenever $\rho^\star(t+1) \geq \rho^\star(t)$. Using (25) and the complementary slackness conditions (22), we also conclude that if $y(t) > 0$ for all $t \in \mathcal{T} = \{\tau_1, \tau_1 + 1, \ldots, \tau_2\}$, $\rho(t)$ is then strictly increasing in $t$ over $\mathcal{T}$.

Combing the results obtained above, we conclude the following only possible dynamics of the optimal control, and the detailed analysis can be found in Appendix D.

(i) $x^\star(0) > z^\star(0)$ when $y_0 > 0$, and $x^\star$ is non-increasing, $y^\star$ decreasing, $z^\star$ non-decreasing in $t$. Either $t_{x=z} = t_{y=0} \leq t_f$ (i.e., $x^\star(t) = z^\star(t)$ and $y^\star(t) = 0$ simultaneously) and from then on the control and the state maintain fixed until $t = t_f$, or $x^\star(t) > z^\star(t)$ and $y^\star(t) > 0$ until the time horizon.

(ii) $x^\star(0) = z^\star(0)$ when $y_0 = 0$, and the control and the state maintain fixed until $t = t_f$.

In addition, once $x^\star(t) = z^\star(t) = u$, we have $u = [\min\{C, D, \tilde{u}\}]^+$ where $\tilde{u}$ is such that $V'(\tilde{u}) = P'(\tilde{u})$.

To conclude, we sketch the dynamics of the optimal control and the state of case (i) in Figure 1. Note that the change of $x^\star$, $y^\star$ and $z^\star$ are not necessarily linear.

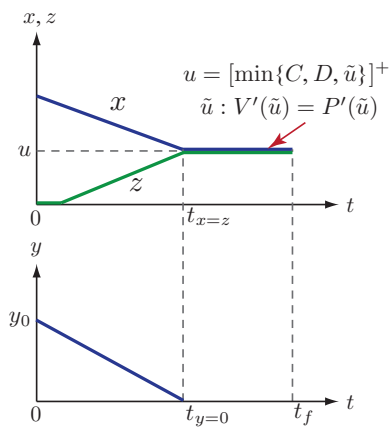

Fig. 1. The optimal policy for a single user

## VI. A GREEDY APPROACH

In this section, we consider a greedy approach to approximating the solution to the SYSTEM problem. The need for a greedy approach can be motivated by the following considerations:

1) depending on the definition of time scale of a step, the convergence time may exceed the time step;

2) some parameters such as $B$, $C$ and $D$ may be stochastic within the time horizon, and we can only observe their realizations a posteriori.

Thus, we consider the following sequence of single-step optimization problems as approximations to the SYSTEM problem. We denote each by SYS_GREEDY $(V, B, C_t, D; t)$,

$$\text{maximize} \quad \sum_{i=1}^{n} V_i(x_i(t)) \tag{28}$$

$$\text{subject to} \quad \sum_{i=1}^{n} x_i(t) \leq C_t \tag{29}$$

$$x_i(t) + \sum_{s=0}^{t-1} x_i^\star(s) \leq D_i \tag{30}$$

$$x_i(t) \geq 0, i = 1, 2, \ldots, n \tag{31}$$

where $(x_i^\star(s), i = 1, \ldots, n)$ is the optimal solution to SYS_GREEDY $(V, B, C_s, D; s)$. Note that because of the myopic nature of each single-step optimization, it does not consider utilizing battery to store energy for future use, and thus the dimension of the greedy formulation reduces to that of the space of $x$.

There are a number of interesting problems to investigate. One of them is whether there is a speed advantage of SYS_GREEDY over SYSTEM, and the other is the optimality gap between them. We will present numerical results on the convergence rate of SYS_GREEDY in Section VII, and in the rest of this section we will characterize the solution of SYS_GREEDY and subsequently the optimality gap.

In the following presentation including Section VII, we limit our discussion to the case where the utility functions of users at a given time step are identical, i.e., $V_i = V$. We denote by $x_i^\star$ and $x_i^{g,\star}$ the optimal solutions of SYSTEM and SYS_GREEDY, respectively. Define $\chi_i^{\star,g} = \mathbf{1}^T x_i^{g,\star}$ and $\chi_i^\star = \mathbf{1}^T x_i^\star$, and denote by $\overline{\chi}_i$, $\mathcal{N}_0$ and $\mathcal{N}_1$ the output of the following procedure.

**Procedure** GetChi(B, C, D)
**Initialize** $\mathcal{N}_0 \leftarrow \{1, 2, \ldots, n\}$, $\mathcal{N}_1 \leftarrow \emptyset$, prev_$\mathcal{N}_0 \leftarrow \mathcal{N}_0$
**repeat**
  prev_$\mathcal{N}_0 \leftarrow \mathcal{N}_0$
  **for** $i \in \mathcal{N}_0$ **do**
    **if** $(\mathbf{1}^T C - \sum_{j \in \mathcal{N}_1} D_j)/|\mathcal{N}_0| < D_i$ **then**
      $\overline{\chi}_i \leftarrow (\mathbf{1}^T C - \sum_{j \in \mathcal{N}_1} D_j)/|\mathcal{N}_0|$
    **else**
      $\overline{\chi}_i \leftarrow D_i$
      $\mathcal{N}_0 \leftarrow \mathcal{N}_0 - \{i\}$, $\mathcal{N}_1 \leftarrow \mathcal{N}_1 \cup \{i\}$
    **end if**
  **end for**
**until** $\mathcal{N}_0 =$ prev_$\mathcal{N}_0$ or $\mathcal{N}_0 \leftarrow \emptyset$
**return** $\overline{\chi}_i, i = 1, \ldots, n, \mathcal{N}_0, \mathcal{N}_1$

The above procedure can be literally described as a best-effort balanced water-filling process. That is, it attempts to fill each water tank (user) the same amount of water (energy), subject to their capacity (demand) constraints, and whenever smaller ones of them overflow, the excess goes to larger tanks in equal amounts. The following proposition asserts that this procedure in fact results in the optimal total per-user amount of energy.

*Proposition 4:* $\chi_i^{\star,g} = \chi_i^\star = \overline{\chi}_i$ for all $i$.

*Proof:* That $\chi_i^{\star,g} = \overline{\chi}_i$ for all $i$ is straightforward, and we prove that $\chi_i^\star = \overline{\chi}_i$ for all $i$ in Appendix C. ∎

*Remark 2:* The balanced property of optimal solutions is a consequence of the concavity and the user-independence of utility function that we assumed. For user-dependent utility functions, we conjecture that the solution may be in a scaled form of balanced power distribution.

Let $p^\star$ and $p^{g,\star}$ be the optimal value of SYSTEM and the sum of optimal values of SYS_GREEDY, respectively. Note that $x_i(t) = \chi_i^\star\left(\frac{C_t}{\mathbf{1}^T C}\right), i = 1, \ldots, n, t = 0, \ldots t_f$, are feasible solutions to the sequence of greedy optimization. The optimality gap is then upper bounded by

$$p^\star - p^{g,\star} \leq (t_f + 1) \sum_{i=1}^{n} V\left(\frac{\chi_i^\star}{t_f + 1}\right) - \sum_{t=0}^{t_f} \sum_{i=1}^{n} V\left(\chi_i^\star \frac{C_t}{\mathbf{1}^T C}\right). \tag{32}$$

Eqn. (32) suggests that the optimality gap would shrink as the power generation capacity $C$ is more balanced among time steps. Moreover, observe that the loss of optimality of the greedy approach results from the myopic optimization that overlooks the use of energy buffer (i.e., battery) to balance energy dispatch between steps. We thus conjecture that the vector $C$ in a roughly ascending order could yield a smaller optimality gap.

## VII. Numerical Illustration

In this section, we illustrate the demand sensitivity to spot market price that we studied in Section IV, using two sample problems that are specified in Table I. We also empirically show the advantage in the convergence rate of the proposed greedy approach compared to the SYSTEM problem, when the respective optimization is performed using the same distributed algorithm based on gradient projection.

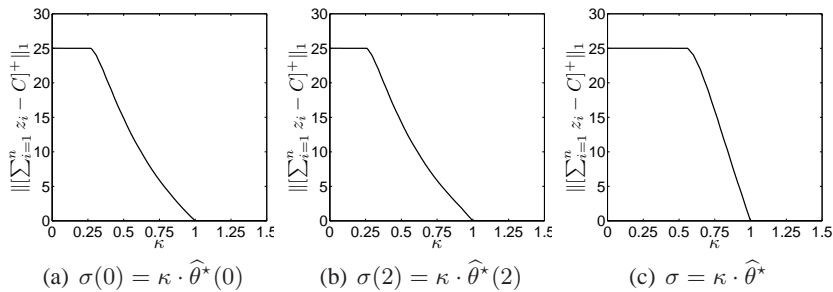

Fig. 2. Demand sensitivity to spot market price for the sample problem $n = 10, t_f = 4$.

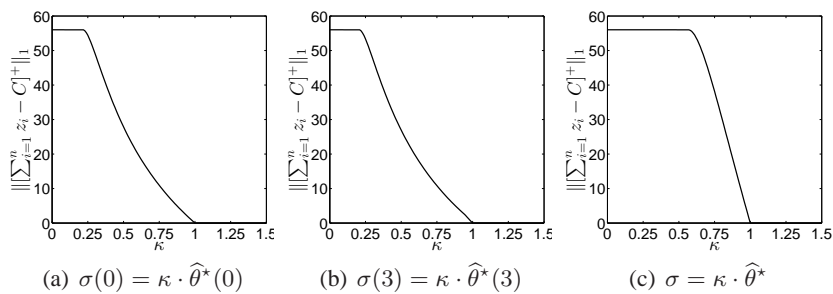

Fig. 3. Demand sensitivity to spot market price for the sample problem $n = 20, t_f = 6$.

As shown in Proposition 2, $\widehat{\theta}^\star$ assumes the threshold value of spot market price $\sigma$, and in Figure 2 and 3 we tune $\sigma$ in three different ways to demonstrate the behaviors of optimal demand. The tests are performed for both sample problems and we elaborate the one with $n = 10, t_f = 4$ for example. In the first scenario, we set $\sigma(t) = \widehat{\theta}^\star(t)$ for $t > 0$ and $\sigma(0) = \kappa \cdot \widehat{\theta}^\star(0)$, where $\kappa$ is a tunable parameter. We scale the third entry of $\sigma$ in the second scenario and the entire price vector in the third one. In Figure 2 and 3, we plot the curves of $\|[\sum_{i=1}^n z_i - C]^+\|_1$ versus $\kappa$, respectively. As can be seen in the figures, the demand increases as $\kappa$ decreases when $\kappa < 1$ until hitting the demand constraint, and remains unchanged when $\kappa \geq 1$, as suggested by Proposition 2.

TABLE I
PARAMETERS OF THE TWO SAMPLE PROBLEMS IN SECTION VII.

|  | $B$ | $C$ | $D$ |
|---|---|---|---|
| $n = 10, t_f = 4$ | $[10, \ldots, 10]^T$ | $[8, 7, \ldots, 4]^T$ | $[1, 2, \ldots, 10]^T$ |
| $n = 20, t_f = 6$ | $[10, \ldots, 10]^T$ | $[10, 9, \ldots, 4]^T$ | $[0.5, 1, \ldots, 10]^T$ |
| $V(a) = \log(1+a)$ and $m(b) = \frac{1}{2}b^2$ | | | |

TABLE II
TOTAL RUNTIME (ITERATIONS) OF SAMPLE PROBLEMS.

|  | SYSTEM | SYS_GREEDY |
|---|---|---|
| $n = 10, t_f = 4$ | 25.0959s(100) | 0.0206s(210) |
| $n = 20, t_f = 6$ | 68.0778s(158) | 0.0316s(197) |

In Table II, we report the respective convergence rates of SYSTEM and SYS_GREEDY. Though these records are machine-dependent, it demonstrates the significant comparative advantage of the latter in speed, and the greedy approach can be used as a good approximation to SYSTEM when the optimality gap bounded by (32) is acceptable.

## VIII. A Distributed Primal-Dual Algorithm: Newton's Method

In this section, we present a distributed second-order algorithm, namely Newton's method, to solve the SYSTEM problem. We first reformulate SYSTEM and approximate it with an equality-constrained problem, and apply Newton's method on the primal problem. Theory and application of distributed Newton's method in network optimization can be found in [17], [18], [19] and etc. Instead of solving the dual problem where updating Lagrange dual variables can be interpreted as dynamic pricing, this algorithm works on the primal side and the solving process can be regarded as dynamic resource slicing [5]. In this section, we assume that $-U_i$ is self-concordant [20] for all $i$ for the proof of convergence. A convex function $f : \mathbf{R} \to \mathbf{R}$ is self-concordant if $|f'''(x)| \leq 2f''(x)^{3/2}$ for all $x \in \mathbf{dom}\, f$, and self-concordant functions include linear, convex quadratic functions, negative logarithm and etc.

We first introduce non-negative slack variables to three inequality constraints, that is,

$$\mathbf{1}^T x_i + p_i = D_i, \quad y_i + q_i = B_i \mathbf{1}, \quad \sum_{i=1}^n z_i + r = C,$$

and using the logarithmic barrier function for non-negative constraints, we approximate our original problem with an equality constrained problem as follows. Let $p = \text{vect}(p_1, \ldots, p_n)$ and $q = \text{vect}(q_1, \ldots, q_n)$. To simplify our notation, we also define $w = \text{vect}(x, y, z, p, q, r)$, that is, for instance, $w_i = x_i$ for $1 \leq i \leq n$ and $w_i(t) = x_i(t)$ and etc. We also define the logarithm for vector $u = (u_1, \ldots, u_m) \in \mathbf{R}^m$ to be

$$\log u = \sum_t \log u_t.$$

Let
$$f(w) = -\sum_{i=1}^{n} U_i(w_i) + \sum_{i=n+1}^{2n} M(w_i) - \tau \sum_{i=1}^{5n+1} \log w_i,$$

and $f$ is also self-concordant for $\tau \geq 1$. Define

$$A = \begin{bmatrix} \Xi & 0 & 0 & I_n & 0 & 0 \\ 0 & I_{n(t_f+1)} & 0 & 0 & I_{n(t_f+1)} & 0 \\ \Gamma & I_{n(t_f+1)} & -\Gamma & 0 & 0 & 0 \\ 0 & 0 & S & 0 & 0 & I_{t_f+1} \end{bmatrix}$$

where $I(m)$ is the $m \times m$ identity matrix,

$$\Xi = \begin{bmatrix} \mathbf{1}^T & & \\ & \ddots & \\ & & \mathbf{1}^T \end{bmatrix}, \Gamma = \begin{bmatrix} H & & \\ & \ddots & \\ & & H \end{bmatrix},$$

and $S = [I_{t_f+1}, \ldots, I_{t_f+1}]$, and also define $b = \text{vect}(D, B_1\mathbf{1}, \ldots, B_n\mathbf{1}, \mathbf{0}, C)$. The original problem can be then approximated as

$$\begin{aligned} \text{minimize} \quad & f(w) \\ \text{subject to} \quad & Aw = b \end{aligned}$$

It can be shown that as $\tau$ approaches zero, the optimum of the above approximating problem converges to that of the original one.

Like the gradient projection method, the Newton's method solves the above problem through an iterated update procedure. In each iteration, we have
$$w^{k+1} = w^k + \gamma^m \alpha^k \Delta w^k,$$

where $\alpha^k$ is the step-size, $\Delta w^k$ is the Newton direction, $\gamma$ is the discounting factor and
$$m = \min\{m' \in \mathbf{Z}_+ \mid w^k + \gamma^{m'} \alpha^k \Delta w^k \succeq 0\}.$$

Let $H^k = \nabla^2 f(w^k)$ and $g_k = \nabla f(x^k)$. The Newton direction is then given by the solution to the following KKT system,
$$\begin{bmatrix} H_k & A^T \\ A & \mathbf{0} \end{bmatrix} \begin{bmatrix} \Delta w^k \\ v^k \end{bmatrix} = -\begin{bmatrix} g_k \\ \mathbf{0} \end{bmatrix}$$

where $v^k$ are dual variables associated with the equality constraint and $\mathbf{0}$ is null matrices with proper dimensions, and it yields
$$\Delta w^k = -H_k^{-1}(g_k + A^T v^k),$$
$$(AH_k^{-1}A^T)v^k = -AH_k^{-1}g_k.$$

Note that $f$ is separable for each $w_i(t)$, and hence $g_k$ only depends on local information with terms

$$\frac{\partial f}{\partial w_i(t)} = \begin{cases} -V'_i(x_i(t)) - \frac{\tau}{x_i(t)}, & w_i(t) = x_i(t) \\ m'(y_i(t)) - \frac{\tau}{y_i(t)}, & w_i(t) = y_i(t) \\ -\frac{\tau}{u(t)} & w_i(t) = u(t), \\ & u = z_i, p_i, q_i, r \end{cases}$$

and $H_k$ is a diagonal matrix with terms on the diagonal

$$\frac{\partial^2 f}{\partial w_i(t)^2} = \begin{cases} -V''_i(x_i(t)) + \frac{\tau}{x_i(t)^2}, & w_i(t) = x_i(t) \\ m''_i(y_i(t)) + \frac{\tau}{y_i(t)^2}, & w_i(t) = y_i(t) \\ \frac{\tau}{u(t)^2} & w_i(t) = u(t), \\ & u = z_i, p_i, q_i, r \end{cases}$$

Therefore, $\Delta w^k$ can be computed from collected data distributedly given the vector $v^k$. Define the Newton decrement $\lambda(w^k)$ as

$$\lambda(w^k) = (\Delta w_k^T H_k \Delta w_k)^{1/2},$$

and use the step-size [18]

$$\alpha^k = \begin{cases} \frac{c}{\lambda(x^k)+1}, & \text{if } \lambda(x^k) \geq \frac{1}{4} \\ 1, & \text{o.w.} \end{cases}$$

where $c$ is some positive scalar that satisfies $5/6 < c < 1$. The description of algorithm is then as follows.

---

**Second-Order Algorithm for Power Distribution**

**User $i$'s algorithm** At iterations $k = 1, 2, \ldots$,
1) Receives from the grid $\Delta w^k$ and $\lambda(w^k)$, and updates $\alpha^k$ and $w^{k+1}$.
2) Computes the associated terms in $g_{k+1}$ and $H_{k+1}$, and communicates them to the grid.

**Grid's algorithm** At iterations $k = 1, 2, \ldots$,
1) Receives from each user $g_k$ and $H_k$.
2) Computes $v^k$, $\Delta w_k$ and $\lambda(w^k)$, and broadcast them to all users.

---

Note that our implementation is different from that in [17] and [18], where the computation of $v^k$ is centralized at the grid, due to our introduction of partial computational capacity of the grid. The self-concordance based convergence analysis is similar to [18] where $\tau$ is assumed greater than 1 to preserve the self-concordance of $f$, and the analysis is omitted for brevity[2].

## IX. CONCLUDING REMARKS

In this paper, we have studied key features of optimal dynamic prices in smart-grid networks. Moreover, an attempt of adopting distributed Newton's method to the model in this paper, namely the SYSTEM problem, is also presented, which is however mainly designed to solve the primal problem and is categorized in the domain of dynamic resource slicing instead of dynamic pricing. In this work, the explicit control structure for time-invariant power production and energy buffering costs has been studied, and continuation of this effort under more general cost functions is one direction of our future research. On the other hand, deviating from the deterministic setting and investigating the optimal pricing with stochastic demand and supply is also one ongoing research project.

---

[2]For the simplicity of presentation, we implicitly assume that the error in computing $v^k$ is negligible. In [18], the authors assume the $H_k$-quadratic norm of error is upper bounded.

# APPENDIX A
# A DUAL-DECOMPOSITION BASED DISTRIBUTED ALGORITHM

Let
$$Q(\nu,\theta) = \sum_{i=1}^n \left(W_i^\star(\nu_i) - R_i^\star(\nu_i)\right) + \theta^T C,$$

and
$$\mathcal{E} = \left\{\text{vect}(\mu,\nu,\theta) \,\middle|\, \theta - H^T\nu_i \succeq 0, \theta \succeq 0, \forall i\right\}.$$

Define $x_i^\star(\nu_i)$ and $y_i^\star(\nu_i)$[3] as the respective unique maximizers of (15) and (16). Let $Q^k = Q(\nu^k, \theta^k)$. We solve DUAL_SYS using the gradient projection method, and the updating rule in the $k$th iteration is given by

$$\widetilde{\nu}_i^{k+1} = \nu_i^k - \alpha^k \nabla_{\nu_i} Q^k = \nu_i^k + \alpha^k\left(Hx_i^\star(\nu_i^k) + y^\star(\nu_i^k)\right)$$
$$\widetilde{\theta}^{k+1} = \theta^k - \alpha^k \nabla_\theta Q^k = \theta^k - \alpha^k C$$

and
$$\text{vect}(\nu^{k+1}, \theta^{k+1}) = \left[\text{vect}(\widetilde{\nu}^{k+1}, \widetilde{\theta}^{k+1})\right]_\mathcal{E},$$

where $\alpha^k$ is a diminishing step-size satisfying

$$\sum_{k=1}^\infty \alpha^k = \infty, \sum_{k=1}^\infty (\alpha^k)^2 < \infty,$$

e.g., $\alpha^k = \frac{1+\Gamma}{k+\Gamma}$ with $\Gamma$ a fixed non-negative number, and $[x_0]_\mathcal{X}$ is the projection of $x_0$ onto a set $\mathcal{X}$, or defined as the following constrained quadratic program

$$\text{minimize} \quad \frac{1}{2}\|x - x_0\|_2^2$$
$$\text{subject to} \quad x \in \mathcal{X}$$

Because all gradients are bounded with diminishing step-size $\alpha^k$, the gradient projection method converges for the dual problem [21], [22], [5]. Since $x_i^\star(\nu_i)$ and $y_i^\star(\nu_i)$ are the respective unique maximizers of (15) and (16) and the strong duality holds, $x_i^\star(\nu_i^\star)$ and $y_i^\star(\nu_i^\star)$ are primal optimal, where $\nu_i^\star$ are the solutions of the dual problem. To summarize, we have the following:

**First-Order Algorithm for Power Distribution**

**User $i$'s algorithm** At iterations $k = 1, 2, \ldots$,
1) Receives from the grid the dual variables $\nu_i^k$
2) Computes $x_i^\star(\nu_i^k)$ and $y_i^\star(\nu_i^k)$, and communicates them to the grid

**Grid's algorithm** At iterations $k = 1, 2, \ldots$,
1) Receives from each user $x_i^\star(\nu_i^k)$ and $y_i^\star(\nu_i^k)$ for all $i$
2) Computes the dual variable $\nu_i^{k+1}$ and $\theta^{k+1}$ for all $i$, and communicates them accordingly to each user

---

[3]With a slight abuse of notation, we will use $x(t)$ to denote the $t$th entry of $x$, and $x(\nu)$ as a function of $\nu$.

The algorithm would terminate when the following stopping criterion is satisfied:

$$\|\text{vect}(\nu^k, \theta^k) - \text{vect}(\nu^{k-1}, \theta^{k-1})\| \leq \epsilon,$$

where $\epsilon$ is a predefined constant.

# APPENDIX B
## PROOF OF PROPOSITION 1

We prove by contradiction. Assume that there exists at least one $i$ such that $(H^T \nu_i^\star)_t = \sum_{s=t}^{t_f} \nu_i^\star(s) < 0$ for some $t$. Let $I = \arg\min_t (H^T \nu_i^\star)_t$. We first note that $x^\star \neq 0$ and further more there exists $t_0 \in I$ such that $x_i^\star(t_0) > 0$. Otherwise, one can switch the value of $x_i^\star(t_0)$ with that of some positive entry in $x_i^\star$ and obtain an improved value of $W_i(x_i; \nu_i^\star)$, which is a contradiction to the fact that $x_i^\star$ is the maximizer of $W_i(x_i; \nu_i^\star)$. Therefore,

1) if there exists $t < t_0$ such that $(H^T \nu_i^\star)_t > (H^T \nu_i^\star)_{t_0}$, let $t_1 = \max\{t : (H^T \nu_i^\star)_t > (H^T \nu_i^\star)_{t_0}, t < t_0\}$ and we then have $\nu_i^\star(t_1) > 0$; otherwise $(H^T \nu_i^\star)_{t_1} < (H^T \nu_i^\star)_{t_0}$. Moreover, $\nu_i^\star(t) = 0$ for $t_1 < t < t_0$ if any. We then set

$$\nu_i^\star(t_0) \leftarrow \nu_i^\star(t_0) + \epsilon$$

and

$$\nu_i^\star(t_1) \leftarrow \nu_i^\star(t_1) - \epsilon$$

where $0 < \epsilon < \min\{\nu_i^\star(t_1), -(H^T \nu_i^\star)_{t_0}\}$, where "$y \leftarrow x$" signifies assigning the value of $x$ to $y$. By doing so, the value of $W_i$ is strictly decreased while that of $R_i$ is non-decreasing and the feasibility constraints are maintained. We thus obtain an improved dual objective value, which leads to a contradiction.

2) If $(H^T \nu_i^\star)_t = (H^T \nu_i^\star)_{t_0}$ for all $t < t_0$ if any, we set

$$\nu_i^\star(t_0) \leftarrow \nu_i^\star(t_0) + \epsilon$$

where $0 < \epsilon < -(H^T \nu_i^\star)_{t_0}$, and the same argument applies as in 1).

# APPENDIX C
## PROOF OF PROPOSITION 4

Note first that $\sum_{i=1}^n \chi_i^\star = \sum_{i=1}^n \overline{\chi}_i = \min\{\mathbf{1}^T C, \mathbf{1}^T D\}$. The proof of this claim is straightforward and we thus omit it here for brevity. Without loss of generality, we assume that $D$ is sorted in an ascending order, and therefore $\overline{\chi}_i$ is non-decreasing in $i$ as the procedure suggests. Moreover, if $\overline{\chi}_j < \overline{\chi}_i$ for some $j < i$, then $\overline{\chi}_j = D_j$. Assume for contradiction that $\chi_i^\star < \overline{\chi}_i \leq D_i$ for some $i$, and there then exists at least one $j \neq i$ such that $\chi_j^\star > \overline{\chi}_j$. Thus,

1) if $j < i$, then $\overline{\chi}_j \leq \overline{\chi}_i$. If $\overline{\chi}_j < \overline{\chi}_i$, we have $\chi_j^\star > \overline{\chi}_j = D_j$, which is infeasible. Therefore, $\overline{\chi}_j = \overline{\chi}_i$ and consequently $\chi_j^\star > \chi_i^\star$. Hence, there exists some $t$ such that $x_j^\star(t) > x_i^\star(t)$. Consider an $\epsilon$ such that $0 < \epsilon \leq \min\{D_i - \chi_i^\star, (x_j^\star(t) - x_i^\star(t))/2\}$ and set $x_j^\star(t) \leftarrow x_j^\star(t) - \epsilon$ and $x_i^\star(t) \leftarrow x_i^\star(t) + \epsilon$. We then obtain a feasible but improved objective value, which is a contradiction.

2) If $j > i$, then $\overline{\chi}_j \geq \overline{\chi}_i$. Hence $\chi_j^\star > \chi_i^\star$ and the same argument applies.

The above argument also applies symmetrically to the case when $\overline{\chi}_i < \chi_i^\star \leq D_i$ for some $i$. Therefore, $\chi_i^\star = \overline{\chi}_i$ for all $i$.

## APPENDIX D
## DYNAMICS OF THE SINGLE-USER OPTIMAL CONTROL

We rule out the suboptimal cases as follows.

1. If $x^\star(0) < z^\star(0)$, then $y^\star(1) > 0$ and $\rho^\star(1) = \rho^\star(0) + M'(y^\star(1)) + \mu^\star(1) > \rho^\star(0)$. Hence, $z^\star(1) \geq z^\star(0)$ and $x^\star(1) \leq x^\star(0)$, and the battery keeps charging until either fully charged before the time horizon and maintain the state, or the time horizon is reached; however, this is clearly suboptimal.

2. If $x^\star(0) > z^\star(0)$, which is only feasible when $y_0 > 0$, let us consider the sequence of events $x^\star(t) \leq z^\star(t)$ and $y^\star(t) = 0$. Let $t_{x \leq z}$ be the first time step when $x^\star(t) \leq z^\star(t)$ or be infinity if it does not happen before the horizon, and similarly we define other quantities.

   2a) If $x^\star(t) \leq z^\star(t)$ happens before $y^\star(t) = 0$, i.e, $\tilde{t} := t_{x \leq z} < t_{y=0}$, then $y^\star(\tilde{t}+1) \geq y^\star(\tilde{t}) > 0$ and thus $\rho^\star(\tilde{t}+1) = \rho^\star(\tilde{t}) + M'(y^\star(\tilde{t}+1)) + \mu^\star(\tilde{t}+1) > \rho^\star(\tilde{t})$. Hence, $z^\star(\tilde{t}+1) \geq z^\star(\tilde{t})$ and $x^\star(\tilde{t}+1) \leq x^\star(\tilde{t})$, and again we are in case 1, which is suboptimal. For the boundary case that $t_{x<z} = t_f$, the suboptimality is also immediate, and we omit the discussion of boundary cases in the following.

   2b) If $t_{x \leq z} > t_{y=0} =: \tilde{t}$, i.e., $y^\star(\tilde{t}) = 0$ while $x^\star(\tilde{t}) > z^\star(\tilde{t})$, this however is infeasible.

   2c) If $t_{x<z} = t_{y=0} =: \tilde{t}$, then $y^\star(\tilde{t}+1) > 0$ and $\rho^\star(\tilde{t}+1) > \rho^\star(\tilde{t})$. Hence, $x^\star(\tilde{t}+1) < z^\star(\tilde{t}+1)$, and we are in case (2a), which is suboptimal.

3. If $x^\star(0) = z^\star(0)$ and $y_0 > 0$, then this is clearly suboptimal.